# How Many Assay Probes to Find One Ore-bearing Asteroid?

Martin Elvis[a,#] and Thomas Esty[b]

*a. Harvard-Smithsonian Center for Astrophysics, USA*

*b. Harvard University, USA*

## Abstract

The number of ore-bearing asteroids could well be small and remote telescopic techniques are inadequate to identify such asteroids confidently. Finding an asteroid that can be profitably mined requires proximate observations from assay probes. Here we use a simple statistical approach to estimate the number of assay probes, $N_{assay}$, needed to find at least one ore-bearing asteroid at a high confidence (90%, 95%, 99%). We present results for a wide range of values of the probability of an asteroid being rich in the resource of interest, $P_{rich}$. We find that $N_{assay}$ depends strongly on $P_{rich}$, for likely values of $P_{rich}$ (<0.5). For a plausible value of $P_{rich} \sim 0.1$ then to obtain 90% confidence that at least one ore-bearing asteroid is found, $N_{assay}$ = 22, and for 99% confidence $N_{assay}$ = 44. A factor two increase in $P_{rich}$ roughly halves $N_{assay}$, while even for $P_{rich} \sim 0.5$, $N_{assay}$ (90%) = 4. Hence any improvement in asteroid characterization prior to sending probes to its proximity would be an effective way to cost-effectively search for valuable resources among the asteroids. Some possibilities for doing so are briefly discussed.

## 1. Introduction

Mining the asteroids for valuable resources is now moving from a conceptual stage (e.g. Lewis et al. 1993) into a development stage, given the formation of companies with this aim[1]. While there are about ten million near-Earth asteroids larger than 20 m diameter (Brown et al. 2013), the number that are rich in the resources of interest is much smaller. Elvis (2013) factorizes the question of how many ore-bearing asteroids there are into four factors: (1) $P_{acc}$, the probability that an asteroid is accessible to current rocketry; (2) $P_{type}$, the probability that the asteroid is of a promising telescopic spectral type; (3) $P_{eng}$ the probability that the engineering involved in the resource extraction if feasible; and (4) $P_{rich}$, the probability that the asteroid is rich in the resource being sought. (I.e. has a sufficiently high concentration of that resource.) Prospecting for these resources is a complex process (Elvis 2013a)

---

[#] corresponding author.

[1] E.g. Planetary Resources URL: http://www.planetaryresources.com (accessed 30 September 2013); Deep Space Industries URL: http://deepspaceindustries.com (accessed 30 September 2013).

Combining these factors Elvis (2013b) estimates that 1/1100 asteroids have high concentrations of water, 1/2000 have high concentrations of platinum group metals (PGMs) and are accessible. These are the two main resources discussed for early asteroid mining. Telescopic observations can improve the odds of finding an asteroid of the most promising type for each resource, based on spectral classification (Bus et al. 2004, deMeo et al. 2009), but the range of resource concentration within a spectral type is still large: four orders of magnitude for PGMs and 2 orders of magnitude for water (Elvis 2013b). For PGMs, Elvis (2013b) estimates $P_{rich} \sim 0.5$ and for water $P_{rich} \sim 0.25$. These appear to be promisingly high values. However these values are highly uncertain. Also, $P_{rich}$(PGM) is only so high for the extremely rare ($P_{type} \sim 0.01$) M-class asteroids (Binzel et al. 2004). Other potential PGM sources, e.g. ordinary LL chondrites and CH carbonaceous chondrites (Lodders and Fegley 1998) are likely to have lower values. Only a spacecraft in proximity to or in contact with an asteroid can make a reliable assay of its resource concentration.

Here we estimate how many such assay probes need to be dispatched to individual asteroids in order to have a high confidence that one of them will be ore-bearing. Our analysis is purely statistical and so does not depend on the nature of the resource once $P_{rich}$ is known. The analysis assumes that each probe operates independently. This is likely to be the case, as waiting for the results of the first probe before sending the second would stretch out the prospecting phase over decades, which is too long for commercial enterprises. A parallel, "swarm", approach is required (e.g. Elvis et al. 2012, Garcia-Yarnoz et al. 2013). Existing estimates of $P_{rich}$ are highly uncertain. Our analysis makes clear that reducing these uncertainties is a priority for prospective asteroid miners. The terrestrial mining industry definition of ore-bearing is not merely that they be sufficiently rich in the resource to be profitably mined, but that the mining can be done profitably (e.g. Sonter 1997). This adds extraction and transport complications that are beyond the scope of this paper, but do not affect the statistical result.

The limitations on our telescopic capabilities for both asteroid composition and mass can be summarized as follows: We can measure the spectra of asteroids in the optical and near-infrared (~0.5 – 2.5 microns). These spectra give indications of the mineral composition of the asteroid surfaces and so tell us whether particular asteroids are carbonaceous, stony or metallic. However, the 24 spectroscopic sub-classes of asteroid give little indication of resource richness (DeMeo et al. 2009). Meteorites show that within these classes the range of resource richness varies over a wide range. For example, PGMs range over 4 orders of magnitude in abundance even among nickel-iron meteorites (Scott, Wasson and Buchwald, 1973; newer papers only report average values), while the amount of water in carbonaceous meteorites, as both hydrated minerals and ice, spans at least a factor of 100 (Jarosewich 1990).

The masses of asteroids are also poorly known, and this obviously affects any estimate of the total value of a resource they contain. Apart from a small fraction of dynamical masses (Hilton 2002) we have to estimate asteroid sizes, and hence volumes, from their brightness at a known distance and illumination angle. In the optical this gives an absolute magnitude[2], H. The albedo of an asteroid (i.e. the fraction of incident sunlight that they reflect), is low and varies by a factor ~5 (~0.05 - ~0.25). Hence the derived volume is uncertain by a factor $5^{3/2}$ = 11, if only reflected light is measured. Knowing the spectroscopic sub-class of an asteroid reduces the uncertainty in albedo somewhat (Thomas et al. 2011, Mainzer et al. 2012) to a 1 sigma uncertainty of a roughly a factor 2. A thermal infrared measurement is more accurate, though technologically harder to obtain. Thermal infrared volumes are good to ~10% (Masiero et al. 2011). Radar observations also give accurate sizes (Ostro et al. 2002), but can only be made for the small fraction of near-Earth asteroids that come within ~0.1 AU of Earth[3].

To get a mass for an asteroid also requires a density. Density depends on composition, for which spectra help. However porosity – the empty fraction within the asteroid – can be large 0.1 – 0.5 and possibly up to 0.7 (Baer, Chesley and Matson 2011, Carry 2012). Hence another factor 2 - 3 uncertainty in mass arises from density. Most asteroid masses are thus uncertain by a factor ~30 with no spectra, or ~6 with spectra.

The two main cases of interest for resource extraction are precious metals (the PGMs) for sale on Earth at ~US$50k/kg, and water, which could be sold in space for similar prices to be used for life support and radiation shielding, or for dissociation into hydrogen and oxygen for rocket fuel. Other materials of interest for in-space sale and use are nickel-iron and gravel, for construction and radiation shielding, respectively. There may also be novel materials found only in asteroids (Elvis and Zeng 2013). The distribution of PGM or water resource concentration is typically inferred from meteorite studies. The threshold value of resource concentration and total mass required to make a mining venture profitable is a commercial decision that the mining company must make. Once this threshold is defined however, the value of $P_{rich}$ follows from the properties of the asteroids.

Regardless of the resource of interest, the number of assay probes, $N_{assay}$, is determined by $P_{rich}$, the probability that the concentration of that resource in an asteroid of interest is high enough to make the asteroid ore-bearing, and by the degree of confidence required by the investors. We use binomial statistics to estimate $N_{assay}$ given $P_{rich}$.

## 2. Binomial Statistics

To find one asteroid in the top 10th percentile of the resource we are seeking one might naively expect that we must make local investigations of about 10 asteroids.

---

[2] An asteroid's absolute magnitude H is the visual magnitude an observer would record if the asteroid were placed 1 Astronomical Unit (AU) away, and 1 AU from the Sun and at a zero phase angle (http://neo.jpl.nasa.gov/glossary/h.html). Conversion from H to an approximate diameter is given at http://neo.jpl.nasa.gov/glossary/h.html. H=22 corresponds to a diameter between 110 m and 240 m for typical albedos.

[3] http://echo.jpl.nasa.gov/~lance/snr/far_asnr18.gif

But we could be unlucky. More formally, if we have a 1 in N chance of finding an asteroid rich enough in our chosen resource, then the probability of finding one on any given trial is $p=1/N$. This problem is equivalent to the problem of selecting a black ball by reaching blindly into a bag containing 1 black ball for every N white balls. This probability is governed by the binomial distribution (e.g. Bevington and Robinson 1992, Chapter 2, Eq. 2.4).

The binomial distribution for the probability of obtaining exactly $x$ occurrences of black in $n$ trials, $P_B$, when the probability of getting one black ball is $p$, is given by

$$P_B(x\,;\,n\,,\,p) = (n/x)\,p^x\,q^{n-x} = [n!/\{x!(n-x)!\}]\,p^x\,(1-p)^{n-x},$$

Where $q = 1 - p$. We would be happy to get more than one black ball, or resource-rich asteroid, so we need to sum over all $x \geq 1$. A simple code lets us determine what value of $n$ we need for a given $p$ for different probabilities of success $P_{x+}$ (i.e. $x \geq 1$). Table 1 gives $n$ for several interesting values of $p$ and $P_{x+}$.

In the asteroid mining case, $n$ is the number of assay missions needed to find at least one asteroid in the required resource richness range, and $p$ is $P_{rich}$, the probability that a given asteroid has the desired resource richness. Figure 1 shows a plot of $n$ vs. $p$ for a wider range of values. Note that the y-axis is logarithmic.

Table 1: Number of assay probes, $N_{assay}$, needed to find one good asteroid as a function of $P_{rich}$ and required confidence level.

| $p$ | n ($N_{assay}$) at given confidence | | |
|---|---|---|---|
| ($P_{rich}$) | 90% | 95% | 99% |
| 0.02 | 114 | 149 | 228 |
| 0.05 | 45 | 59 | 90 |
| 0.1 | 22 | 29 | 44 |
| 0.15 | 15 | 19 | 29 |
| 0.2 | 11 | 14 | 21 |
| 0.25 | 9 | 11 | 17 |
| 0.3 | 7 | 9 | 13 |
| 0.4 | 5 | 6 | 10 |
| 0.35 | 6 | 7 | 11 |
| 0.45 | 4 | 6 | 8 |
| 0.5 | 4 | 5 | 7 |

The result is quite different from the naïve expectation that 10 trials is enough if $p=0.1$. The binomial distribution probability of finding at least one in the top 10th percentile with ten trials is 65%. To reach 90% we must make 22 trials. Forty-four trials are needed to reach 99% probability. A positive result is that there is a 66% to 94% probability there will also be a second good asteroid if there is a first.

Even at quite high values of $P_{rich}$ the assay probe numbers are not all that small. At $P_{rich}$ = 0.5, appropriate for finding PGMs in the rare M-class asteroids (Elvis 2013), $N_{assay}$ = 4 – 7, depending on the confidence required. At $P_{rich}$ = 0.25, appropriate for carbonaceous asteroids that may be rich in water (Elvis 2013), $N_{assay}$ = 9 – 17, again depending on the confidence required.

Figure 1: Number of assay probes needed to find at least one resource-rich asteroid at 90%, 95% and 99% confidence, as a function of the probability that an asteroid exceeds an ore-bearing threshold. The integer steps are inherent to binomial statistics.

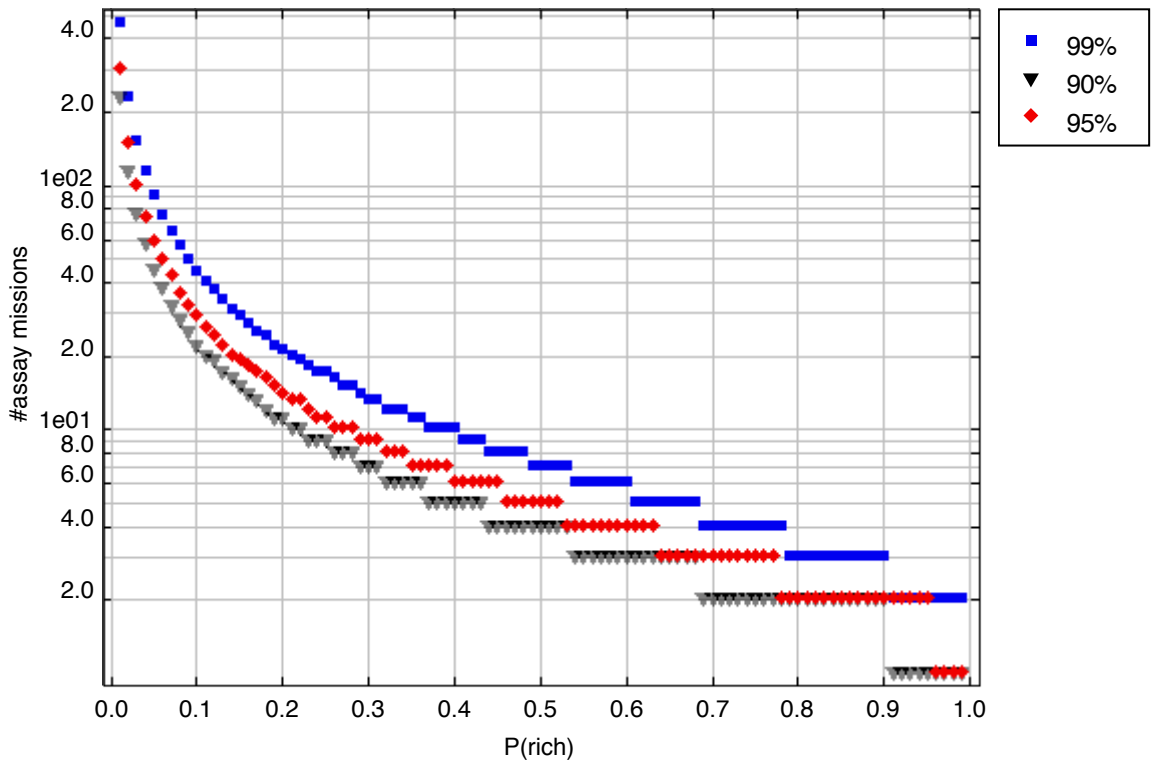

## 3. Discussion

The number of assay probes, $N_{assay}$, is important for keeping the costs of finding rich asteroids manageable. It is clear from Table 1 and Figure 1 that the difference between $P_{rich}$ = 0.1 and $P_{rich}$ = 0.5 is almost an order of magnitude in $N_{assay}$. At 90% confidence $N_{assay}$ decreases from 22 to 4, and at 99% confidence decreases from 44 to 7, probes respectively. Even a smaller improvement in $P_{rich}$, from 0.1 to 0.2 halves $N_{assay}$, from 22 to 11 (90% confidence), which could be economically significant.

Such large $N_{assay}$ values mean that even cheap probes can add up to a significant cost for a mining venture. (Some numbers are presented below, Sec. 3.1 and 3.2.) Hence reducing $N_{assay}$ by investing in improving $P_{rich}$ is likely to be cost effective. This implies that significant effort should be focused on pushing existing telescopic methods to their observational limit so as to increase $P_{rich}$ to as high a value as possible. Some possibilities are discussed in Sec. 3.3.

### 3.1 Assay Probe Cost, Mass and Volume

If $N_{assay} \sim$ dozens then the methods to be employed in looking for asteroid mining candidates are quite constrained. Because the transit time from one good asteroid candidate to the next is likely to be a year or so, serial investigations candidate NEOs by one spacecraft will take decades, which is too long for a commercial venture. Instead we must consider a parallel approach using a fleet, or swarm, of smaller spacecraft to investigate a number of NEOs simultaneously. A swarm approach requires low mass, volume and cost per assay probe.

Neither mass nor payload volume are prohibitive. As an example, a Falcon-9 launcher can launch a 4.8 mt payload to Geostationary Transfer Orbit (GTO http://www.spacex.com/falcon9), or roughly 1.7 mt to a NEO Hohmann (1960) transfer orbit (scaling naïvely using delta-v $\sim$2.5 km s$^{-1}$ GTO-NEO vs. 2.8 km s$^{-1}$ LEO-GTO, http://www.lr.tudelft.nl/?id=29271&L=1&id=29271). If we allocate 1 mt to a carrier for the $N_{assay}$ probes, then each probe can have a mass of 700/$N_{assay}$ kg, or 29 kg to 58 kg for 12 – 24 probes. This is a factor 2 – 4 larger mass/probe than is being considered by Planetary Resources for the Arkyd series, which have advertised masses of ~12 kg (http://www.planetaryresources.com/2013/06/planetary-resources-prepares-for-launch-of-the-arkyd-series-of-spacecraft/). The Falcon 9 fairing volume is unrestrictive at over 100 m$^3$ (Space Exploration Technologies 2009), > 4 m$^3$– 8 m$^3$/probe.

The cost of obtaining the assay information needs to be a modest fraction of the value of the resources ultimately retrieved, perhaps 10%. For a nominal value of ~US$1.2 B[4]. That fraction would cap the *in situ* prospecting cost at ~US$120 M. To launch a SpaceX Falcon 9 currently costs ~US$56 M[5]. The cost of each probe, including mission operations, must then be ~US$64 M/$N_{assay}$ ~US$2.5 – 5 M, for 12 – 24 probes. This comparable to the cost per probe discussed by Planetary Resources and Deep Space Industries, two asteroid mining companies, perhaps following the same reasoning as laid out here. Simultaneous operation of dozens of probes will require mostly autonomous commanding, decision-making and data receipt, if cost is to be contained. Any increase in cost may require larger or more ore-rich asteroids to be assayed and mined in order to make a profit. However these rapidly become scarce (Stuart and Binzel 2004, Elvis 2013).

### 3.2 Assay Probe Instrumentation for Resource Richness Determination

To keep each probe small and cheap will require a minimal instrument package. Radio tracking to determine mass, and optical imaging to determine shape and mean density are essentials. This instrumentation will surely be included on the Arkyd and Firefly probes, from Planetary Resources and Deep Space Industries respectively. However neither technique can determine resource richness.

---

[4] The approximate value of a 100 m diameter PGM-rich NEO (Elvis 2013b).

[5] http://www.spacex.com/falcon9

To determine resource richness requires knowing the elements present in the asteroid and their abundances. An X-ray imaging spectrometer can determine these well. X-ray emission lines provide much more specific information than the broad mineral features seen in optical-infrared spectra (e.g. Dunn et al. 2010, Thomas and Binzel 2010). The X-ray flux from the Sun produces fluorescence in each surface element to produce distinctive emission lines, giving the elemental composition of the surface in ten or more elements from oxygen through iron in the ~0.5 – ~8 kiloelectronvolt (keV) energy range, when compared with the input Solar flux (Allen et al. 2013). Adding an instrument like REXIS on OSIRIS-REx would require ~5 kg and ~0.01 m$^3$. This would stay within the parameters derived in Sec. 3.1, though several months of exposure would be needed to map the NEO surface (Allen et al. 2013).

Surface composition may not, however, be a good guide to bulk composition. Space weathering can coat the asteroid surface with nano-phase iron (Pieters et al. 2000), which will mask the true composition. Sublimation from the top several centimeters of water-bearing asteroids may give a misleading assay of the interior composition. In these cases some means of sub-surface probing will be needed. Laser ablation/drilling with optical/infrared spectroscopy to determine the composition of the ejected gas (Gibbings et al. 2013) could avoid both having the probe make contact with the asteroid and the complications involved in mechanical sub-surface digging (e.g. Daniels 2013).

### 3.3 Improving Remote P$_{rich}$ Estimates

As N$_{assay}$ is highly sensitive to P$_{rich}$, efforts to increase P$_{rich}$ even modestly could have a high payoff. Improved studies of meteorites and their connections to asteroid types are thus worth exploring.

Surprisingly little detailed composition data is available for meteorite families. Haak and McCoy (2003) report mean richness values (in µg/g or % weight) for nine elements (Re, Ir, Ni, Co, Cu, Au, Ga, Ge, S) for iron and stony-iron meteorites, but the original data have not been collated and listed, but remain scattered in the meteoritical literature. Creating an SQL-searchable database of all meteorite data would be a first step. If the data remain too sparse to produce useful distributions of potential resource concentrations for each meteorite type, then a systematic analysis of all meteorite falls would be rewarding to undertake.

Even with good meteorite type resource distributions in hand, there remains the problem of connecting asteroid types determined telescopically with these meteorite types (Burbine et al. 2002). The single asteroid 2008 TC3, which was studied briefly before its impact with Earth produced a few kilograms of meteorites recovered from the Sudanese desert (Jenniskens et al. 2009), may point the way. Roughly one meter-sized asteroid per month collides with Earth (Brown et al.

2002). These asteroids move fast across the sky and require high cadence sensitive surveys to find them with even a few days advance notice. The newly NASA-funded ATLAS project ("Asteroid Terrestrial-impact Last Alert System", Jedicke et al. 2012), and other upgraded NEO surveys, may provide a supply of these "death plunge" asteroids. If so then direct asteroid-meteorite comparisons may become routine. Higher $P_{rich}$ values will result.

## 4. Summary and Conclusions

We have shown how many *in situ* assay probes are needed to find at least one resource-rich asteroid with high confidence. For likely values of the probability of a given asteroid being sufficiently resource-rich, $P_{rich}$, the number of probes, $N_{assay}$, lies in the dozens. $N_{assay}$ is highly sensitive to $P_{rich}$. A factor two increase in $P_{rich}$ roughly halves $N_{assay}$, a considerable saving. Encouragingly, there is a 66% to 94% probability there will also be a second good asteroid if there is a first.

For $N_{assay}$ = 24 the probes must be of modest mass (<29 kg), less modest volume (~ 4 m$^3$), but especially low cost (<US$2.5M). The Arkyd and Firefly probes (being developed by Planetary Resources and Deep Space Industries, respectively) fit this profile.

However, to determine resource concentrations from an *in situ* probe requires knowing the bulk composition more accurately than optical-near-infrared spectroscopy can determine. *In situ* imaging X-ray spectroscopy in the few kiloelectronvolt (keV) energy range can determine surface abundances for a dozen or so elements in a few months.

If the surface is unrepresentative of the bulk composition, however, some sub-surface method must be used. Laser ablation coupled with remote spectroscopy may suffice. If not then the complexities of establishing contact with the asteroid and digging beneath its surface must be faced. Keeping within the design constraints will then be challenging.

Improvements in our knowledge of meteorite resource-richness could help increase $P_{rich}$. This could begin with construction of a modern database of existing measurements. A program targeted at creating good composition distributions for large numbers of meteorites in each class could be a valuable follow-on. Direct asteroid-meteorite comparisons would greatly enhance the prospecting of NEOs and so reduce the need for large numbers of *in situ* assay probes.

The results given in this paper, combined with the modest numbers of likely ore-bearing NEOs found by Elvis (2013b), show that NEO characterization prior to *in situ* probes can be an effective way to advance the search for valuable resources among the asteroids.

## Acknowledgements

We thank Doug Finkbeiner, Branden Allen, Jae-Sub Hong, Karen Daniels, Alison Gibbings, and Jonathan McDowell for helpful discussions. ME thanks the Aspen Center for Physics, where this paper was completed.## References

Allen B.T., et al., 2013, *The REgolith X-Ray Imaging Spectrometer (REXIS) for OSIRIS-REx: Identifying regional elemental enrichment on asteroids*, SPIE, in press.

Baer J., Chesley S.R., and Matson R.D., 2011, *Astrometric Masses of 26 Asteroids and Observations of Asteroid Porosity*, Astrophysical Journal, 141, 143.

Bevington P.R. and Robinson K., 1992, *Data Reduction and Error Analysis for the Physical Sciences*, [McGraw-Hill, Boston], ISBN 0-07-911243-9.

Binzel R.P., et al., 2004, *Observed spectral properties of near-Earth objects: results for population distribution, source regions, and space weathering processes*, Icarus, 170, 259 – 294.

Brown P., Spalding R.E., ReVelle D.O., Tagliaferri E. & Worden S.P., 2002, *The flux of small near-Earth objects colliding with the Earth*, Nature, 420, 294-296.

Brown, P.G., et al., *A 500-kiloton airburst over Chelyabinsk and an enhanced hazard from small impactors*, 2013, Nature, Volume 503, Issue 7475, pp. 238-241

Burbine T., McCoy T.J. Meibom A., Gladman B., & Keil K., 2002, *Meteoritic Parent Bodies: Their Number and Identification*, in "Asteroids III", eds. W.F. Bottke Jr., A. Cellino, P. Paolicchi and R.P. Binzel, [U. Arizona and LPI, Tucson], ISBN 0-8165-2281-2, pp. 653-667.

Bus, S.J., Binzel R.P., and Burbine T.H., 2000, *A New Generation of Asteroid Taxonomy*, Meteoritics & Planetary Science, vol. 35, Supplement, p.A36

Carry B., 2012, *Density of Asteroids*, Planetary and Space Science, 73, 98 – 118.

Daniels, K., 2013a, *Rubble-pile Near Earth Objects: Insights from Granular Physics*, in "Asteroids: Prospective Energy and Material Resources", ed. Viorel Badescu, [Springer, Heidelberg], Chapter 11.

DeMeo F.E., Binzel R.P., Slivan S.M. & Bus S.J., 2009, *An extension of the Bus asteroid taxonomy into the near-infrared*, Icarus, 202, 160–180.

Dunn T.L., McCoy T.J., Sunshine J.M., McSween H.Y. Jr., 2010, *A Co-ordinated Mineralogical, Spectral and Compositional Study of Ordinary Chondrites: Implications for Asteroid Spectroscopic Classification*, LPI, 41, 1750.

Elvis M., Kasper J., Landau D., Lantoine G., Marrese-Reading C., Mueller J., Russell R.P., Strange N., Ziemer J.K., Nash A., Yeomans D., 2012, *A Swarm of NEO Micro-satellites and Robotic Precursors*, Global Space Exploration Conference, GLEX-2012.03.3.1x12307.

Elvis M., 2013a, *Prospecting Asteroid Resources*, in "Asteroids: Prospective Energy and Material Resources", ed. Viorel Badescu, [Springer, Heidelberg], Chapter 4, pp. 81 - 130.

Elvis M., 2013b, *How Many Ore-Bearing Near-Earth Asteroids?*, Planetary and Space Science, in press.